# Evaluating the Impact of Missing Data Imputation through the use of the Random Forest Algorithm

Adam Pantanowitz and Tshilidzi Marwala
School of Electrical and Information Engineering, University of the Witwatersrand, Private Bag x3,
Wits, 2050, Republic of South Africa

*Abstract*—This paper presents an impact assessment for the imputation of missing data. The data set used is HIV Seroprevalence data from an antenatal clinic study survey performed in 2001. Data imputation is performed through five methods: Random Forests, Autoassociative Neural Networks with Genetic Algorithms, Autoassociative Neuro-Fuzzy configurations, and two Random Forest and Neural Network based hybrids. Results indicate that Random Forests are superior in imputing missing data in terms both of accuracy and of computation time, with accuracy increases of up to 32% on average for certain variables when compared with autoassociative networks. While the hybrid systems have significant promise, they are hindered by their Neural Network components. The imputed data is used to test for impact in three ways: through statistical analysis, HIV status classification and through probability prediction with Logistic Regression. Results indicate that these methods are fairly immune to imputed data, and that the impact is not highly significant, with linear correlations of 96% between HIV probability prediction and a set of two imputed variables using the logistic regression analysis.

## I. INTRODUCTION

MISSING data are a common difficulty encountered in many real-world situations and studies, and creates difficulties with data analysis, study and visualization [1], [2]. The missing information also reduces insight into the data, and the underlying cause for the fact that data are missing may make the data of particular interest. Furthermore, decision policies made by a decision making system often cannot exact a decision without all the information at hand. For this reason, it is important to find effective and viable methods of imputing data, and furthermore, the effect of this imputation should be considered such that insight is gained into the validity of decisions made by such decision making systems. The problem is assessed in the context of a real-world data set taken from an HIV sero-prevalence survey performed in South Africa in 2001 [3].

This paper evaluates the concept, classification, problem and treatment of missing data. Background into the various methods and paradigms used in the paper are then considered, followed by a description into the implementation of these concepts. The data set is considered, and thereafter feature selection on the data is described. Comparisons are drawn in the paradigms used, and the impact and sensitivity analysis is performed. Finally, a discussion is presented and conclusions are drawn.

## I. MISSING DATA

Missing data are a problem inherent and common in data collection, especially when dealing with large, real-world data sets. Missing data are a problem since statistical methods have difficulty in performing when data are unknown. Studies have highlighted the need to research decision support systems when key information is missing or inaccessible [4]. The effect of missing data on such decision support systems is marked, and it is shown that results are degraded by simply assigning an arbitrary value to the missing data elements.

In the context of surveys, missing data may result for a number of reasons. Incomplete variable collection from subjects, non-response from subjects declining to provide information, poorly defined surveys, or data being removed for reasons such as confidentiality are some of the explanations for missing data [1], [2].

### A. Categorization of Missing Data and the Missing Data Mechanism

Missing data can be categorized based on the pattern of missingness and the missingness mechanism. The methods with which the missing data are dealt are dependent on the category into which the data fall. Three broad categories for pattern missingness are defined: monotone missingness, file matching, and general missingness [5], [6]. If a set of variables for a given instance are $y_1; ...; y_k$, monotone missingness occurs if when a missing value $y_j$ occurs, the variables can be ordered such that $y_{j+1}; ...; y_k$ are also missing. The pattern of file matching occurs when two variables are never jointly observed. Arbitrary missingness is a missingness pattern which occurs when neither of the former two patterns applies.

Missing data are often classified into one of three mechanisms, as defined by Little and Rubin [5]. The mechanisms are listed as follows in order from least to most dependent on other information.

1) Missing Completely At Random (MCAR) arises if the probability of a missing value is unrelated to the variable value itself or any other variable in the data set.

2) Missing At Random (MAR) arises if the probability of missing data of a variable could depend on other variables in the data set, but not on the variable's own value.
3) Non ignorable case in which the probability of missing data is related to the value of the variable even if other variables are known/controlled.

In the MCAR case, data cannot be predicted using any information in the set, known or unknown. For the MAR mechanism, there is a correlation between the missing data and the observed data, but not necessarily on the value of the missing data [7].

*B. Dealing with Missing Data*

A number of strategies have been devised for dealing with missing data. The simplest means is discarding the instances in which missing data occur (a complete-case method), which is both inefficient and leads to potentially biased conclusions and observations. This is also not practical if a large proportion of data are missing. This method leads to information waste as information is discarded [1]. Despite this, the method is used commonly in practice [2]. Other techniques include *available-case procedures*, *weighting procedures* and *imputation-based procedures* [7]. The latter is discussed further here, since imputation methods can be applied to the MCAR and MAR cases [8].

Imputation techniques involve predicting the values of the data which are missing. Two categories of techniques exist, model-based techniques and non-model based techniques. Non-model based approaches include mean imputation and hot-deck imputation. These techniques have been said to decrease the variance estimates in statistical procedures [7]. Furthermore, such techniques may result in standard errors and bias on results. Model-based approaches include regression-based techniques, Expectation Maximization [9] and Multiple Imputation [8]. Neural network based approaches have been successfully implemented a number of times [1], [9], [10].

### III. BACKGROUND

*A. Random Forests*

Ensemble or Network Committees are algorithms in machine learning which combine individual paradigms to form combinations which are often more accurate than the individual classifier alone [11]. In the classification case, overall predictions can be obtained from such a network using a weighted or an un-weighted voting system; in the regression case, overall predictions can be chosen through an averaging technique. Obtaining a general understanding of why such methods succeed is an active area of research [11], [12].

A *Decision Tree* is a tree with nodes which contain information corresponding to attributes in the input vectors. This information is used to follow a decision path for a given set of input attributes, depending on either thresholding nodes (as in the case of a continuous variable) or categorical nodes (as in the case of categorical data) [13].

Even though decision trees have appeal for being straightforward and fast, they are prone to being overly adapted to the training data or to a loss in accuracy for generalization through tree pruning [14].

"Random Forest" (RF) is an algorithm which generalizes ensembles of Decision Trees [15] through bagging (Bootstrap Aggregation) which combines multiple random predictors in order to aggregate predictions [16]. They allow for complexity without over-generalizing the training data [14]. RF can be used for both regression and classification, and has been used with success in the context of missing data [13]. Random Forests were first introduced in 2000 by Breiman, and "Random Forests" is a trademark of Cutler and Breiman [11]. Each tree in the RF is grown according to algorithm 1, and each tree forms an independent member of the forest [13].

---
**Algorithm 1** The growing of a tree in a RF [15]
1) Select the splitting criterion (constant for the RF), $m$ to be much less than the number of input variables, $M$ ($m \ll M$)
2) For $N$ Training Samples, sample $N$ cases at random *with replacement*
3) Grow each tree with the $N$ sampled cases:
    a) $m$ variables are selected at random from $M$ and the best split on these $m$ is used to split at each node
    b) Grow each tree as much as possible (without *pruning*)

---

If, as stated in [16], [13], $\Theta$ is the possible variables, and $h(x, \Theta)$ denotes a tree grown using $\Theta$ to classify a vector $x$, then a RF can be defined as

$$f = \{h(x, \Theta)\}, k = 1,2,..., K, \qquad (1)$$

in which $\Theta_k \subseteq \Theta$. Thus, each tree in the forest contains an individually selected subset of the overall collection of attributes.

The error rate of the RF is shown to be dependent on two properties [15]:

- The *correlation* between any two trees in the forest
- The *strength* of an individual tree in the forest

The *correlation* refers to how similar one tree is to another, and increasing the correlation between trees increases the Forest Error Rate (FER). The *strength* refers to how strong a classifier the tree is, and increasing the strength of individual trees decreases the FER. The parameter $m$ is directly proportional both to *correlation* and to *strength*, so there is an optimal range of $m$ at which the *correlation* is minimized and the *strength* is maximized.

Sample with replacement results in some of the training set not being used in training (approximately a third of the training data) [15]. These data are referred to as "out-of-bag" (oob) data that are used to get an unbiased estimate of the performance of the RF, which is unlike *cross-validation* which may be biased [17]. Furthermore, oob data are used in predicting variable importance, which is discussed further in section VI. Information regarding strength and correlation can also be obtained from the oob methods, allowing one to gain insight into the forest [17]. The *proximity* is an $N \times N$ matrix obtained by running all the data down the tree, and if two cases are in the same terminal node, their proximity is increased by one [15]. This is a useful property which can be used in locating outliers or estimating missing data.

RFs have been an area of active research in the last few years for their numerous advantageous features and high success [11]. RFs are said to work fast, have excellent accuracy offering improvements over single classification and regression trees (CART), be impervious to over-fitting the data, run efficiently on thousands variable numbers (no dimensionality problems), give an unbiased self-assessment and variable importance assessment, and have effective methods for missing data estimation and for outlier location [11], [15]. These properties make the RF algorithm a logical candidate for this missing data study.

*B. Other Paradigms*

For comparative purposes, other learning paradigms and hybrid networks are considered, consisting of elements such as Neuro-Fuzzy (NF) Networks, Multilayer Perceptron (MLP) Neural Networks (NN), and Genetic Algorithms (GA). These are generally connected in Autoassociative configurations [10], and the details are discussed further in section IV. While these are introduced briefly here, the interested reader is encouraged to visit the relevant references for more in-depth detail.

*1) Multilayer Perceptron Neural Network:* MLPs are neural networks which consist of an interconnection of the processing elements, generally placed in three classes: the input layer, the output layer and the hidden layer [18]. A process of supervised learning allows the weights of the network to be adjusted until a satisfactory error is obtained between the output and target comparison, yielding a feed-forward network capable of modeling the complex input output relationships [19]. The Neural Network architecture consists of the selection of the number of nodes (or neurones) in the hidden layer; how many inputs and outputs there are; and the type of activation function used. A number of different optimization strategies are available in training the network, such as conjugate gradient descent [18].

*2) Neuro-Fuzzy:* A fuzzy inference system (FIS) can be developed if we have knowledge expressible in terms of linguistic rules. Fuzzy systems involve interpretation of if-then rules through a process of fuzzification (resolving the antecendent to a degree of membership), fuzzy operation and implication (the consequent assigns a fuzzy set to the output) [20]. Fuzzy inference is the entire process of mapping from a given input to a given output using fuzzy logic. While a fuzzy system makes use of natural language, a NN can be used if we have data for training. Drawbacks of each of the systems are seen to be complementary, and thus the integration of the two systems is logical. The FIS offers an advantage in terms of learning capability, while the extraction and learning of rules is a problem well suited to ANNs [21].

Neuro-Fuzzy systems consist of rule sets and inference systems combined with or governed by a connectionist structure for optimisation and adaptation to given data. Adaptive Neuro Fuzzy Inference System (ANFIS) implements a Takagi Sugeno (TS) FIS and consists of five layers, the first of which is for fuzzification of the input variables [21]. The second layer employs a T-norms operation which computes the rules of the antecedent. The third layer normalizes rule strength, while the fourth layer determines the consequent of the rule. It is important to note that in TS FIS, the consequent part of the rule is mathematically zero order or first order [20]. The fifth layer is the output layer, which computes the weighted global output as a combination of all the incoming information. The schematic architecture of this system is presented in [22], [23]. The system can employ grid partitioning or subtractive clustering techniques [20]. In the learning process, the parameters associated with the membership functions change – this change is an optimisation essentially facilitated by a gradient vector [23]. Using a combination of back-propagation and with the use of a least squares method [20], the fuzzy inference system is able to learn from the model data. A TS system is suited for modelling of non-linear systems by interpolating multiple linear models [20].

*3) Genetic Algorithm:* Biologically motivated modeling techniques have led to the development of stochastic optimization techniques, which are useful in control applications [24]. Genetic Algorithms (GA) are essentially optimisation search methods that are broadly used to solve optimisation problems, especially with large, difficult to interpret sets of data [24]. Genetic algorithms employ their heuristic search by modeling techniques of natural evolution including: crossover; inheritance; mutation; and selection [25].

Through a process of random search, genetic algorithms exploit the properties of biological evolution in order to solve optimisation problems [25]. A global optimum can be found through the process of modelling natural selection: the convergence exists due to the fitness of an individual in a given population dominating over another individual [26]. Each individual represents an element in the search space which may be an appropriate solution to the problem [25]. The individuals thereafter go through a process of evolution, and survival of the fittest ensues.

Initially, individuals are selected in a population at random (within the range of appropriate values for which an optimum will be found) [24]. Through processes of crossover (in which a model of genetic recombination is utilised); inheritance; mutation and selection, individuals compete for reproductive rights and resources [26]. The more dominant genes (or good genes) will propagate through the species – and thus there is eventual convergence.

Two parents may produce offspring which are better suited to the environment than either of the parents. Through this process, offspring become better adapted to the environment through generations of selection, since inferior offspring are eliminated through evaluation of their fitness (their inability to obtain resources and inability to breed) [25].

*4) Autoassociative Networks:* Autoassociative networks are system models in which the model is trained to recall the input. This means that the number of outputs is equal to the number of inputs [10]. The Autoassociative Neural Network Encoder (or *autoencoder*) usually has a smaller number of nodes in the hidden layer than the number of inputs (or outputs). This creates what is referred to as a bottleneck. The autoencoder network can detect missing datum elements $\{x_u\}$ by forward propagating the known elements and a predicted value for the unknown elements, and minimising the overall error between the input and the output. The error is generally quantified, and minimised, using an intelligent search method, such as a genetic algorithm, as in [10] and [9]. The error can be evaluated as follows:

$$E = \left( \begin{Bmatrix} x_u \\ x_k \end{Bmatrix} - f\left( \begin{Bmatrix} x_u \\ x_k \end{Bmatrix}, \{w\} \right) \right)^2, \quad (2)$$

in which $x_u$ indicates the unknown value(s), $x_k$ indicates the known value(s), and the function essentially defines the input-output relationship of the neural network with weights $w$.

## IV. METHODOLOGY AND SYSTEM TOPOLOGIES
### A. General Systems

*1) Random Forest (RF):* A number of different methods and system topologies are investigated. The main topology investigated is a Random Forest (**RF**). The RFs used throughout the analysis generally have 70 trees, the parameter for minimum size of terminal nodes set at 7, and have the variables to be randomly sampled at each split (*m*) set to 3 (since there are 14 inputs (*M*), this is a reasonable number since it is required to be much less than the number of inputs [17]). This combination was determined experimentally to be the optimal set of parameters through maximisation of the number of hits (i.e. the number of correct predictions). Regression RFs are used when predicting ordinal variables, which are encoded to be continuous values ranging from 0 - 1, and Classification RFs when predicting categorical variables, such as HIV Status (as discussed in table I). RF was implemented through a MATLAB interface [27].

Since each RF makes a single prediction, an attempt was made to form a RF to predict each of the fourteen variables, and combine this with a GA to form a type of RF based autoassociative network. A similar procedure is followed for the Neuro-Fuzzy system, and discussed in section IV-A3. This method did not yield favourable results. However, the fourteen forests were used to impute different missing variables, depending on which was missing. In order to achieve this, the methodology presented in figure 1 is employed.

*2) Autoassociative Neural Network:* The Autoassociative Neural Network (implemented in MATLAB using the Netlab toolbox [28]) combined with the GA (implemented using the GAOT toolbox [29]) (AANN-GA) is set up with a bottleneck in the number of hidden nodes. For the 14 input/output systems, the optimal number of hidden nodes, determined experimentally, was set to 11. This allows the data to be generalized while rejecting redundancy. The number of training cycles was determined to be 400, the minimum point of the validation curve. A linear activation function was used with scaled-conjugate descent training.

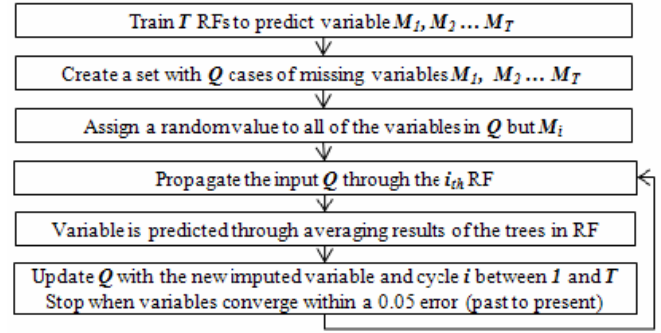

Fig. 1. Flow chart indicating concept used in imputing multiple missing variables with regression and classification RFs

*3) Autoassociative Adaptive Neuro-Fuzzy Inference System:* The Autoassociative Neuro-Fuzzy System with a GA (AANFGA) implements a network of 14 ANFIS networks. Since each network predicts a single value, a system of 14 is set up in order to minimise the error between the input and the output in an autoassociative configuration. Each of the ANFIS systems uses subtractive clustering to train, with a training radius of 0.2, 20 training epochs, and a stopping criterion of 0.01.

### B. Hybrid Systems

*1) Random Forest & Autoassociative Neural Network Hybrid:* In this topology, the RF is placed in 'cascade' with an AANN-GA, yielding what is referred to in this paper as the **RF-AANN-GA**. The RF is used to predict a set of missing variables in an experiment set, and the predictions are recorded. These predictions are then used as limits for the search space of the GA, in the AANN-GA system. Since the variable range is 0 - 1, a tolerance of 10 % is placed on the variable (i.e. 0.05 is added and subtracted from the predicted value) such that the GA has a slightly broader search space. This corresponds approximately to a four year interval for age (for example). The principle is that by limiting the search space, the AANN-GA will be improved in performance. A similar principle is successfully applied in [1], in which C4.5 Decision Trees are used rather than the RF to limit the search space. The results obtained by [1] for the C4.5 AANNGA are also compared to the presented systems. The AANNGA and RF have the same structures and parameters as the aforementioned standalone optimised structures.

*2) Autoassociative Neural Network & Random Forest Hybrid:* In this topology, the AANN-GA system is placed in cascade with a RF, yielding what is referred to in this paper as the **AANN-GA-RF**. The principle behind the operation here is that the RF learns the underlying problems in the AANN-GA system and compensates for them. In order to achieve this hybrid system, the data are divided into four sets: Training; Validation; Testing and Experimental. The training and validation data are used to train and obtain the best model for the AANN-GA using early stopping [30]. Thereafter, data are removed from the testing as well as the experiment set to yield artificially incomplete sets, and these incomplete sets are propagated through the AANN-GA to obtain missing data predictions from the AANN-GA. The testing data and imputed values are made to form a complete set. This testing set is then used as a training set for the RF, with the target being the original, correct data. In this way, the RF is realised to track missing data using the error from the AANN-GA as a reference, and compensate for it. The experimental set is then used to test the RF.

## V. Data Evaluation and Preprocessing

Preprocessing of data is crucial in order for the data to be of appropriate form for the machine learning paradigms. The data set used is based on a National HIV and Syphilis Sero-Prevalence Survey of Women attending antenatal clinics in South Africa, and is taken from the study performed in 2001 [3]. The data consist of survey information from 16 743 pregnant women. The variables contained in the data set are outlined in table I.

TABLE I
OUTLINE OF DATA SET VARIABLES (ADAPTED FROM [31])

| Variable | Abbreviation | Data Type | Range | Variable Type |
|---|---|---|---|---|
| Province (Location) | - | Integer | 1 - 9 | Categorical |
| Age | - | Integer | 12 - 50 | Ordinal |
| Education | Edu | Integer | 0 - 13 | Ordinal |
| Gravidity | Gra | Integer | 1 - 12 | Ordinal |
| Parity | Par | Integer | 0 - 9 | Ordinal |
| Father's Age | FathAge | Integer | 12 - 90 | Ordinal |
| HIV Status | HIV | Binary | 0/1 | Categorical |
| RPR Test Status | RPR | Binary | 0/1 | Categorical |
| Race | - | Integer | 0 - 5 | Categorical |

Note that the data range given is for once the variables have been processed, as is discussed below. Gravidity refers to the number of times a woman has been pregnant, and parity the number of times the woman has given birth. Father's Age refers to the age of the father responsible for the current pregnancy. Education is specified as 0 (no education); 1 - 12 (for grades 1 through to 12); and 13 (tertiary education). Province categorises a person in to one of the 9 South African provinces, and race categorises a person in to one of 6 race categories.

Since we are dealing with a real-world study, involving missing data as discussed in section II, the data contain inherent errors. In order to yield a complete set, the data are first processed according to the following logical rules, and any datum satisfying the following is labeled as "missing":

- Data cannot be negative
- Age of female must be between 12 to 50
- Age of male must be greater than 12
- Gravidity cannot be less than parity
- The education level cannot exceed 13
- All fields must be valid as specified.

The data were normalised, and binary encoding was applied to the categorical data of race and province, since these variables are not ordinal, and this may lead to problems for the learning paradigms [31].

## VI. Feature Selection and Multiple Missing Values

Features of the data can be selected for by the RF algorithm. The RF algorithm provides estimates of the importance of each variable in the data set in predicting a given output [15]. This is performed with the oob data. On predicting for variables in the given data set, it is found, as expected, that variables for which effective estimates are obtained on missing data have high correlations with other variables in the set.

If we are to test the impact of two or more missing variables, there are a large number of permutations that require testing.

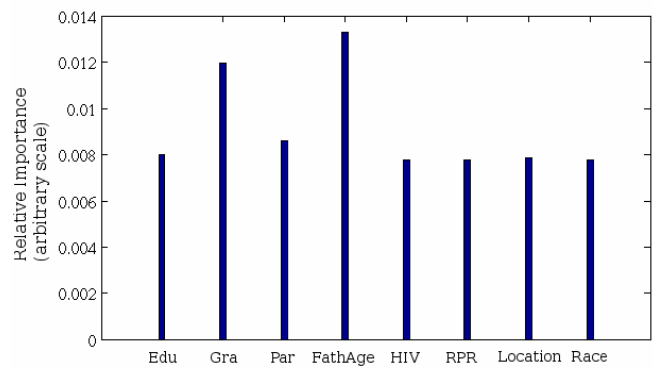

Fig. 2. Graph indicating variable importance for imputing the Age variable

However, feature selection allows us to test combinations of variables which have been selected according to their mutual correlations. This allows for a more meaningful analysis to be performed. This analysis offers insight into why certain variables are certainly removable when predicting other certain variables, and offers valuable information when attempting to overcome the curse of dimensionality [32]. Fortunately, RF is not prone to the curse of dimensionality, but ANFIS unfortunately is. This analysis explains the reason for this, and it is to be noted that this is the reason that fewer tests are done on multiple missing variables. This knowledge is applied in the impact assessment of section VIII.

TABLE II
VARIABLE IMPORTANCE FOR EACH POTENTIALLY MISSING VARIABLE

| Variable | High Importance Variables | Low Importance Variables |
|---|---|---|
| Age | Father's Age; Parity | HIV Status; RPR |
| Education | All approximately of equal importance (Age slightly dominant) | |
| Gravidity | Parity | All others |
| Parity | Gravidity | All others |
| Father's Age | Age | All others |
| HIV Status | Education; Race | All others |
| RPR Status | Province | All others |
| Province | Race; Parity | Age; Education |
| Race | Province | All others |

An example of the feature selection output in terms of variable importance is presented for the age variable in figure 2. Table II presents the notable importance in the other variables for each variable examined. The predictions are generally logically sensible, for example, gravidity should be highly correlated with parity, and age with the father's age should generally be correlated. However, the importance of parity in predicting province (location) seems interesting.

## VII. COMPARISON AND RESULTS

Table III indicates the results on testing the missing data prediction ability of the various systems. The results are found by predicting missing data of the indicated variables, and calculating the percentage of values accurately predicted within the specified ranges. MAR and MCAR are not distinguished, and the Non-Ignorable case is neglected in the analysis. Note that the C4.5 AANN-GA results are obtained for the appropriate ranges from [1]. The ranges are indicated in the table (for example, age prediction is assessed for prediction percentages within 1, 2, 4, 6 and 10 years).

Testing is performed to determine the best of the techniques specified in section IV. It is evident from the result that the RF and RF hybrids outperform the other methods of missing data prediction. There is significant improvement in the RF from the commonly used AANN-GA method, with an average percentage increase of 7.6 % for the indicated categories. Education prediction increases by an average of 31.2 % from the AANN-GA to the RF across the specified categories. The improvement from the AANN-GA to the AANN-GARF is a significant one, indicating that the hybrid method of section IV is working, but the results are comparable to the standalone RF. Furthermore, for the case of the RF-AANNGA it is observable through experimentation that narrowing the search bounds of the GA improves the performance. Thus, introducing the AANN-GA with larger search bands starts to degrade the performance of the hybrid, indicating that this hybrid's results are suffering from problems within in the AANN-GA. The AANN-GA and AANF-GA perform relatively badly in different aspects: age prediction (for the AANF-GA) and education prediction (for the AANN-GA). The RF performs well in all respects and does not suffer drawbacks in either of these categories.

TABLE III
RESULTS OF PERCENTAGE PREDICTION ACCURACY FOR VARIOUS METHODS WITHIN THE SPECIFIED RANGES

| Quantity | Range (Within) | RF | RF-AANN-GA | AANN-GA-RF | AANF-GA | AANN-GA | C4.5, AANN-GA [1] |
|---|---|---|---|---|---|---|---|
| Age (years) | 1 | 41.1 | 35.1 | 40.2 | 22.0 | 34.7 | – |
| | 2 | 62.3 | 55.9 | 60.8 | 36.7 | 54.7 | 52.3 |
| | 4 | 85.7 | 83.0 | 85.0 | 54.0 | 81.1 | 79.4 |
| | 6 | 95.0 | 92.8 | 93.9 | 68.7 | 90.5 | 89.6 |
| | 10 | 99.2 | 98.5 | 99.0 | 80.7 | 96.5 | 97.9 |
| Education (grades) | 0 | 16.7 | 19.7 | 15.4 | 6.5 | 5.5 | – |
| | 1 | 53.5 | 48.1 | 51.5 | 18.5 | 24.3 | 52.1 |
| | 2 | 76.9 | 69.4 | 75.7 | 34.5 | 35.8 | 69.5 |
| | 3 | 88.3 | 83.7 | 88.3 | 46.5 | 46.4 | 79.4 |
| | 5 | 93.1 | 90.8 | 93.2 | 70.0 | 54.8 | 91.8 |
| Gravidity (Instances) | 0 | 88.0 | 88.1 | 88.1 | 0 | 88.0 | 80.4 |
| | 1 | 98.3 | 98.2 | 98.2 | 13.7 | 98.2 | 97.1 |
| | 2 | 99.5 | 99.4 | 99.4 | 35.7 | 99.4 | – |
| | 3 | 99.8 | 99.7 | 99.7 | 67.0 | 99.8 | 99.6 |
| | 5 | 100.0 | 100.0 | 100.0 | 95.0 | 100.0 | 100.0 |
| Parity (Instances) | 0 | 89.4 | 87.6 | 89.5 | 0 | 87.9 | 60.8 |
| | 1 | 98.5 | 98.3 | 98.4 | 21.5 | 98.2 | 92.9 |
| | 2 | 99.6 | 99.5 | 99.6 | 52.0 | 99.4 | – |
| | 3 | 100.0 | 99.9 | 100.0 | 74.0 | 99.8 | 89.6 |
| | 5 | 100.0 | 100.0 | 100.0 | 94.0 | 100.0 | 97.9 |
| Father's Age (years) | 1 | 28.8 | 27.9 | 28.3 | 3.5 | 27.7 | – |
| | 2 | 45.7 | 45.6 | 46.2 | 11.5 | 45.9 | 41.7 |
| | 4 | 74.1 | 72.8 | 73.6 | 22.5 | 72.1 | 68.6 |
| | 6 | 86.3 | 86.2 | 86.7 | 32.0 | 86.1 | 82.7 |
| | 10 | 95.3 | 94.4 | 95.0 | 53.0 | 94.2 | 93.2 |

While the hybrid methods appear to show potential, the computational time trade-off for the use of these methods (due to the need to cascade NNs with GAs) is not warranted for performance improvement. This is especially so in lieu of the relative computation time taken, as indicated in IV. It is to be noted that the study to obtain this table was performed in MATLAB, using the tools specified in section IV. Thus, the programming is not standardised, and this result should be treated as a basic evaluation. That said, it is to be noted that RF is generally documented as being relatively fast machine learning tools [11], [14], [15], [17], and this is clearly reflected in the table.

TABLE IV
RELATIVE COMPUTATION TIME TAKEN FOR THE VARIOUS INDICATED METHODS FOR PROPAGATION THROUGH 5000 INSTANCES WITH MISSING DATA POINTS

| Method | Training Time (s) | Propagation Time (s) |
|---|---|---|
| RF | 0.04 | 0.5 |
| AANF-GA | 797.4 | 50964 |
| AANN-GA | 20.7 | 628.3 |

TABLE V
HIV STATUS PREDICTION CONFUSION MATRIX FOR RF CLASSIFIER RUN ON EXPERIMENT SET

| Confusion Matrix | Predicted Negative | Predicted Positive | Percentage Error (%) |
|---|---|---|---|
| Actual Negative | 2902 | 1732 | 37.4 |
| Actual Positive | 449 | 875 | 33.9 |

The HIV status of the individual is predicted by a RF classifier, and the results are presented in table V. The other configurations were also used to predict HIV status, however, this is not discussed further, since the results of the RFs alone are used in the impact and sensitivity assessment of section VIII. The AANN-GA accuracy obtained at 64.2 % with an F-measure of 0.43. The classification results obtained are lower than those found in [10], and this is a trade-off to be discussed in the recommendations of section IX-B.

## VIII. IMPACT AND SENSITIVITY ASSESMENT

The impact of estimating the missing data is evaluated within this section by evaluating three aspects: the statistical impact on the data, the impact on HIV classification, and the impact on a decision making system. This assessment gives an overall picture, since it offers insight into the effects of imputation within the data (statistical assessment [33]), and on the effects of imputation on classifiers [4], [10] and on a decision making system. Study variables are selected based on their mutual correlations, and based on the prediction performance of the RF predictor. Note that for each missing variable (s), two sets are defined, one which has variables imputed through RFs (*Sets RFx*) and one which has the variable(s) randomly assigned (*Sets Rx*). The randomly defined sets act as an experiment control to ensure that the imputed results presented are not spurious. Note that when these sets are used in conjunction with an HIV classifier or decision making system, as in sections VIII-B and VIII-C, HIV data is not used as an input to impute the missing data. Using the variable selection technique discussed in section VI, and the results of section VII the following sets are defined:

- The original complete target data set (*Set T*)
- A single imputed variable with average prediction performance - Age (*Sets RF1A & R1A*)
- A single imputed variable with poor prediction performance
- Education (*Sets RF1B & R1B*) (important for HIV prediction as per table II)
- A single imputed variable with good prediction performance
- Gravidity (*Sets RF1C & R1C*)
- Two imputed variables which are of high mutual importance (as per table II) - Age and Father's Age (*Sets RF2A & R2A*)
- Three imputed variables - Age, Education, Father's Age (*Set RF3A & R3A*)
- Four imputed variables - Age, Education, Father's Age, Gravidity (*Set RF4A & R4A*).

Some of the evaluation techniques include the goodness of fit measures in terms of the KS test [33] and the Mahalanobis Distance (the mean distance is taken) [34]. These give a statistical measure of the similarity between the data sets, and are regarded as a good measure of the fit between results. The mean squared error offers a relative indication of the difference between data sets, and is calculated as

$$MSE = \frac{1}{N} \sum_{i=1}^{N} (T_i - P_i)^2 , \quad (3)$$

in which $N$ is the entire data set, $T_i$ represents the $i^{th}$ target value and $P_i$ represents the $i^{th}$ predicted value.

### A. Statistical Impact

The statistical impact on the data is measured through a number of statistical measures. Missing data are imputed by the RF algorithm, and this creates a new set which is compared to the original, complete set. A control set which contains a randomly imputed set is created for comparison. This allows one to observe that the variation in values between the imputed values and the true set cannot be attributed to random factors. Tables VI, VII and VIII present statistical results for the variables of age, education and gravidity respectively.

TABLE VI
STATISTICAL IMPACT FOR SINGLE MISSING VARIABLE: AGE

| Measure | Age (years) | | |
|---|---|---|---|
| | Target Set (*T*) | Imputed Set (*RF1A*) | Random Set (*R1A*) |
| Mean | 25.00 | 25.3 | 31.2 |
| 1st Quartile | 20 | 21 | 22 |
| Median | 24 | 25 | 31 |
| 3rd Quartile | 29 | 29 | 41 |
| Standard Deviation | 6.3 | 5.4 | 11.0 |
| Variance | 40.0 | 29.0 | 120.3 |
| Combined MSE | - | 10.5 | 195.5 |
| Mean Mahalanobis Distance | - | 0.73 | 3.96 |
| Linear Correlation (with Target Set) | - | 85.92 % | 2.01 % |
| Maximum Percentage Deviation | - | 84.2 % | 163.2 % |

A Quantile-Quantile Plot [33] (QQ Plot) allows one to view the deviation in the distributions of a given variable. The extent of deviation from a straight line indicates distribution deviation. Figure 3 presents QQ Plots for the real set and the imputed variable and for the real set and randomly imputed values. Note the former plot is fairly linear, with the interpolated line intercepts close to the origin and to the coordinates (1, 1), indicating a good distribution match, whereas the latter plot is fairly non-linear.

TABLE VII
STATISTICAL IMPACT FOR A SINGLE MISSING VARIABLE: EDUCATION

| Measure | Education (grade) | | |
|---|---|---|---|
| | Target Set (Set T) | Imputed Set (Set RF1B) | Random Set (Set R1B) |
| Mean | 9.4 | 9.4 | 6.6 |
| 1st Quartile | 8 | 9 | 3 |
| Median | 10 | 10 | 7 |
| 3rd Quartile | 12 | 10 | 10 |
| Standard Deviation | 2.8 | 1.4 | 3.7 |
| Variance | 7.8 | 1.8 | 14.1 |
| Combined MSE | - | 6.4 | 29.8 |
| Mean Mahalanobis Distance | - | 0.24 | 2.83 |
| Linear Correlation (with Target Set) | - | 42.7 % | 0.4 % |
| Maximum Percentage Deviation | - | 177.7 % | 200 % |

TABLE VIII
STATISTICAL IMPACT FOR A SINGLE MISSING VARIABLE: GRAVIDITY

| Measure | Gravidity (instances) | | |
|---|---|---|---|
| | Target Set (Set T) | Imputed Set (Set RF1C) | Random Set (Set R1C) |
| Mean | 2.1 | 2.0 | 6.5 |
| 1st Quartile | 1 | 1 | 4 |
| Median | 2 | 2 | 6 |
| 3rd Quartile | 3 | 3 | 9 |
| Standard Deviation | 1.4 | 1.3 | 3.2 |
| Variance | 1.8 | 1.6 | 10.1 |
| Combined MSE | - | 0.23 | 30.88 |
| Mean Mahalanobis Distance | - | 0.87 | 16.0 |
| Linear Correlation (with Target Set) | - | 93.8 % | 3.25 % |
| Maximum Percentage Deviation | - | 200.0 % | 127.2 % |

*B. Impact on HIV Classification*

We define for classification the categories in table IX [10]. From the definitions, we can define evaluation metrics in order to evaluate the classifiers. First, the accuracy of the classifier is defined [10]:

$$Accuracy = \frac{TN + TP}{TN + FN + TP + FP}. \quad (4)$$

*Sensitivity* allows one to assess how well the classifier can recognise positive samples, and is measured as $\frac{TP}{TN+FN}$.
*Specificity* measures how well the classifier recognises samples as negative, and is evaluated as $\frac{TN}{TN+FN}$.
*Precision* is a measure of the percentage of samples correctly specified as positive, $\frac{TP}{TP+FP}$ [35]. Note that *Recall* (Re) is the same measure as *Sensitivity* [36]. The $F_{measure}$ is used to assess a system when a single number is preferred [36],

$$F_{measure} = \frac{2 \times P \times R}{P + R}, \quad (5)$$

where $P$ is the precision and $R$ is the recall.

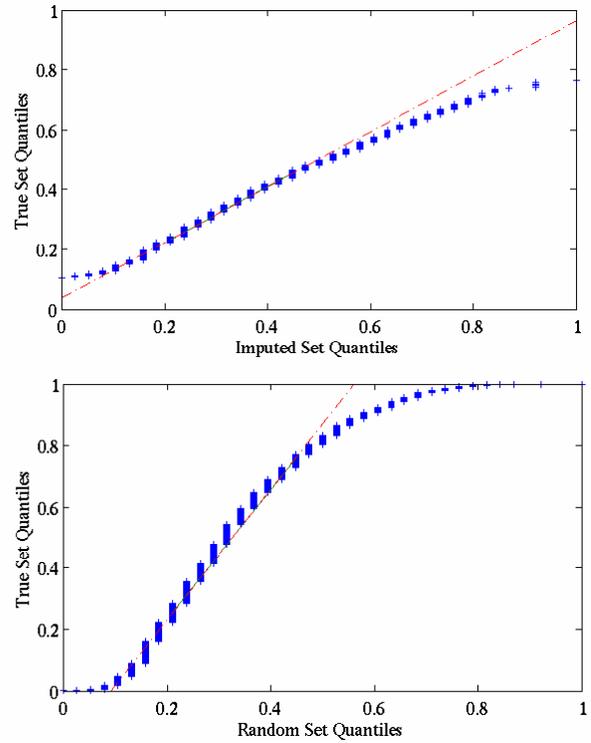

Fig. 3. QQ Plots of the target data set of age with the RF imputed set (above) and with the set with age randomly imputed (below)

TABLE IX
ERRORS FOR BINARY CLASSIFICATION [10]

| | Predicted Negative | Predicted Positive |
|---|---|---|
| **Actual Negative** | True Negative (TN) | False Positive (FP) |
| **Actual Positive** | False Negative (FN) | True Positive (TP) |

In order to assess the impact of imputed data on HIV Classification, a number of imputed data sets and data sets with randomly assigned values are propagated through a RF classifier. Tables X and XI compare these results with that of the classification results of the target (complete) set (presented in table V in a confusion matrix).

TABLE X
HIV CLASSIFICATION RESULTS FOR DATA SETS WITH ONE IMPUTED VARIABLE

| Set | T | RF1A | R1A | RF1B | R1B | RF1C | R2C |
|---|---|---|---|---|---|---|---|
| TP | 837 | 936 | 506 | 926 | 768 | 839 | 310 |
| TN | 2986 | 2787 | 3057 | 2766 | 2463 | 2955 | 4040 |
| FP | 487 | 388 | 1577 | 1868 | 2171 | 1679 | 594 |
| FN | 1648 | 1847 | 818 | 398 | 556 | 485 | 1014 |
| Accuracy (%) | 64.2 | 62.5 | 59.8 | 61.2 | 54.2 | 63.7 | 73.0 |
| Sensitivity (%) | 63.2 | 70.7 | 38.2 | 79.9 | 58.0 | 63.4 | 23.4 |
| Specificity (%) | 85.9 | 87.8 | 78.9 | 87.4 | 81.6 | 85.9 | 79.9 |
| F Measure | 0.44 | 0.46 | 0.29 | 0.45 | 0.36 | 0.44 | 0.28 |

TABLE XI
HIV CLASSIFICATION RESULTS FOR DATA SETS WITH TWO, THREE AND FOUR IMPUTED VARIABLES

| Set | T | RF2A | R2A | RF3A | R3A | RF4A | R4A |
|---|---|---|---|---|---|---|---|
| TP | 837 | 947 | 308 | 1033 | 354 | 1122 | 109 |
| TN | 2986 | 2391 | 3589 | 2133 | 3892 | 1825 | 4358 |
| FP | 487 | 2243 | 1045 | 2501 | 742 | 2809 | 276 |
| FN | 1648 | 377 | 1016 | 291 | 970 | 202 | 1215 |
| Accuracy (%) | 64.2 | 56.0 | 65.4 | 53.1 | 71.3 | 49.5 | 75.0 |
| Sensitivity (%) | 63.2 | 71.5 | 23.3 | 78.0 | 26.7 | 84.7 | 8.2 |
| Specificity (%) | 85.9 | 86.4 | 77.9 | 88.0 | 80.0 | 90.0 | 78.2 |
| F Measure | 0.44 | 0.42 | 0.23 | 0.43 | 0.29 | 0.43 | 0.13 |

Evident from these tables is that the classifier, though indicated to be of average performance from the F measure, shows resilience and almost immunity to the sets with estimated data, especially with 1 or 2 imputed variables. The effects of the random data sets are evident, with F measures dropping into the 0.2 range. This experimental control ensures that the variables do affect the classifier, and thus indicates that the experimental results are not spurious.

*C. Impact on Decision Making System*

A Logistic Regression [37] (LR) decision making system is designed, which, based on input, computes the probability that the output variable belongs to a given set. The output variable is specified to be HIV Status, and we thus obtain probability of an individual's membership to the HIV Positive class or Negative class. The original set T is propagated through the regressor, and a set of probabilities that individuals are HIV positive is obtained. Thereafter, the sets with various imputed variables are propagated through the regressor, to yield a set of probabilities that individuals (with imputed demographics) are HIV positive. The probabilities resulting from the original set (T) are compared with the probabilities resulting from the imputed sets (RF1A, etc.). Results of this test are presented within this section. The probabilities are expressed as percentages, and where tests of fit are involved (e.g. the KS test) the relevant result data are compared with the results from the original set T. The results are fairly similar for the different single imputed variable sets, and thus results presented in table XII are for the statistical differences in the regressor outputs due to set T, and due to the sets with one, two and four imputed variables.

TABLE XII
MEASURES OF THE RESULT OF LOGISTIC REGRESSION ANALYSIS FOR THE ORIGINAL DATA SET WITH SETS WITH ONE, TWO AND FOUR IMPUTED VARIABLE THROUGH COMPARISON

| Set | T | RF1A | R1A | RF2A | R2A | RF4A | R4A |
|---|---|---|---|---|---|---|---|
| 1st Quartile (%) | 18.7 | 18.8 | 19.1 | 19.7 | 19.1 | 20.6 | 23.9 |
| Median (%) | 24.6 | 24.8 | 26.2 | 26.2 | 26.2 | 26.8 | 34.4 |
| 3rd Quartile (%) | 27.9 | 27.9 | 32.8 | 27.9 | 32.5 | 27.9 | 44.1 |
| Mean (%) | 22.6 | 22.6 | 25.3 | 23.1 | 25.3 | 23.4 | 33.4 |
| Variance (%) | 89.7 | 89.8 | 130.8 | 88.5 | 128.1 | 92.4 | 230.4 |
| Linear Correlation (%) | - | 98.9 | 87.7 | 96.4 | 87.3 | 95.3 | 77.8 |
| KS Test | - | 0.02 | 0.21 | 0.12 | 0.21 | 0.18 | 0.49 |
| Mean Squared Error | - | 1.8 | 38.3 | 6.7 | 37.9 | 9.23 | 214.3 |

Once again, the results indicated in table XII predict a fair amount of immunity to imputed data on the probabilities given by the LR analysis. The original data set and the set with one, two and even four imputed values do not significantly change the predictions of the LR. This is emphasized in figure 4 which indicates the QQ plot of the two probability distributions from the logistic regression analysis. First, the prediction from the logistic regression analysis is plotted against the prediction for the 2 imputed variable set. Second, the prediction from the logistic regression analysis is plotted against the prediction for the 4 imputed variable set. Note that the plots are fairly linear, indicating high similarity between probability results from the LR. As the number of imputed variables increases, the correlation decreases. This indicates that the number of imputed variables does indeed have an effect on the results. Note that the randomly generated sets (R1A, R2A and R4A) indicate significant deviation on propagation through the LR analysis. These sets cause the variance in the data to increase by a significant amount. This indicates that the LR analysis is sensitive to the data that are tested, thus validating the experiment.

IX. DISCUSSION

*A. Impact on Society*

Through the study of the impact of imputation of missing data on these types of systems, it is notable that missing data imputation does not significantly negatively impact on classifiers and decision-making systems. A considerable impact on society can be noted from the work. A decision-based system for preliminary HIV classification in a healthcare or study context is invaluable. Furthermore, missing data which renders potentially useful information in a given study meaningless can, through appropriate imputation, be made into meaningful data. This can help studies of this nature to uncover the statistical trends that are said to be discarded through the removal of missing entries (in, say, a complete-case method). Therefore, decisions can be made with some confidence on instances which previously were impossible due to missing information.

matches at terminal nodes. In the context of standardizing testing impact in the way that is presented in this paper, the terminal node method was not feasible. However, this certainly should be investigated in future work.

The RF HIV classifier does not perform exceedingly well when compared with classifiers using the same data [10], and requires investigation. However, for the application, the RF classifier was used for comparative purposes in impact assessment, and thus the performance of this element is not crucial when compared with the RFs used for data imputation.

## X. CONCLUSION

Missing data causes significant information loss in studies as information is wasted, and no insight is gained into the underlying causes for the missing data. Through the use of survey data of results from an HIV sero-prevalence, this paper investigates five machine learning paradigms in order to obtain imputed data for an impact assessment on the effects of missing data: RFs, AANN-GA, AANF-GA, RF-AANN-GA and AANN-GA-RF. From the five, RFs are chosen for impact analysis due to their superiority in both prediction accuracy and in computation time taken. Data sets are generated with one, two, three and four imputed variables each, and these sets are used for evaluating impact. Impact is determined in three ways: through evaluating statistical deviations of the imputed variables relative to the true values; through an HIV classifier performance; and through a logistic regression analysis for probability prediction. Results indicate that these decision making systems are in fact rather immune to the imputation of missing data, when adequate imputation techniques are used. These results imply that decision based systems are therefore able to make informed decisions where previously impossible on instances with missing information through imputation.

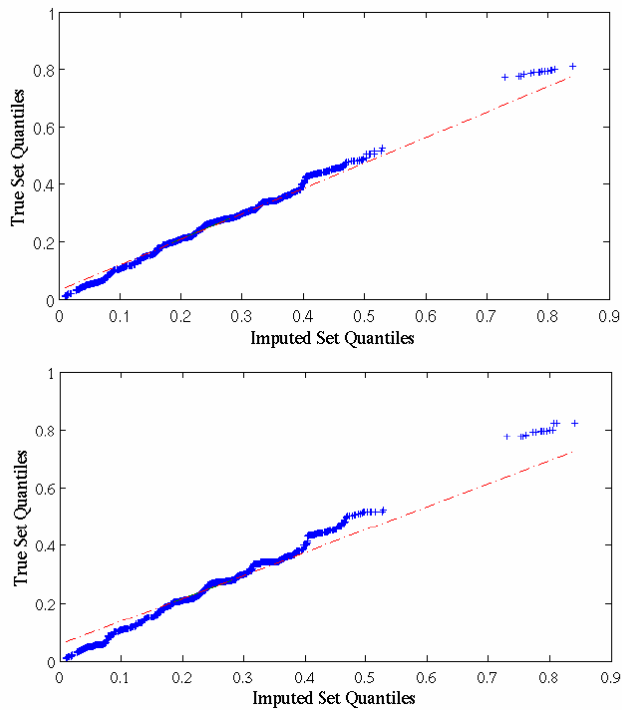

Fig. 4. QQ Plot of result from logistic regression probability analysis on HIV status for the true set (T) and the set with 2 imputed variables (R2A) (above); and logistic regression probability analysis on the HIV status for the true set (T) and the set with 4 imputed variables (R4A) (below)

### B. Recommendation for Future Work

The results obtained from the AANF-GA (which implements ANFIS) are relatively considerably poor. Despite the fact that ANFIS struggles with high dimensionality data, train was possible through subtractive clustering. However, it was not feasible to train the ANFIS system with grid partitioning unless variables were removed, and this would affect the comparison with other systems. In order to maintain a critical study, all learning paradigms were tested with the same number of input variables. This said, the ANFIS system may still have promise and should not be completely disregarded.

The RF implementation within this work is through RF regression and classification. However, as discussed in section III-A, RF has a built in method of assessing missing data by initially estimating data, and thereafter evaluating

REFERENCES

[1] G. Ssali, T. Marwala. "Estimation of missing data using computational intelligence and decision trees." Proceedings of IEEE International Joint Conference On Neural Networks, Hong Kong.
[2] N. J. Horton, K. P. Kleinman. "Much ado about nothing: a comparison of missing data methods and software to fit incomplete data regression models," in The American Statistician, Vol. 61, No. 1, pp. 79, 2007.
[3] Department of Health, South African Government. "Summary report: national HIV and Syphilis sero-prevalence survey of women attending public antenatal clinics in South Africa, 2001." http://www.info.gov.za/. Last accessed 31 July, 2008.
[4] M. K. Markey, G. D. Tourassi, M. Margolis, D. M. DeLong. "Impact of missing data in evaluating artificial neural networks trained on complete data," in Computers in Biology and Medicine 36, pp. 517 - 525, Elsevier, 2006.
[5] R. J. Little, D. B. Rubin. Statistical Analysis with Missing Data. John Wiley & Sons, 2002.
[6] M. L. Ziegler. "Variable selection when confronted with missing data." Ph.D Dissertation, University of Pittsburgh, 2006.
[7] D. J. Fogarty. "Multiple imputation as a missing data approach to reject inference on consumer credit scoring." http://interstat.statjournals.net/YEAR/2006/articles/0609001.pdf. Last accessed 10 August, 2008.
[8] K. H. Yuan, P. M. Bentler. "Three likelihood-based methods for

mean and covariance structure analysis with non-normal missing data," in Sociological Methodology, pp. 165 - 200, 2000.
[9] F. V. Nelwamondo, S. Mohamed, T. Marwala. "Missing data: a of neural network and expectation maximisation techniques," Current Science, Vol. 93, No. 11, pp. 1514 - 1521, 2007.
[10] B. Leke Betechuoh, T. Marwala, T. Tettey. "Autoencoder networks for HIV classification," in Current Science, Vol. 91, No. 11, pp. 1467-1473, 2006.
[11] G. Biau, L. Devroye, G. Lugosi. "Consistency of random forests and other averaging classifiers." Journal of Machine Learning Research, to appear, 2008.
[12] L. Masisi, F. V. Nelwamondo, T. Marwala. "The effect of structural diversity of an ensemble of classifiers on classification accuracy." IASTED International Conference on Modelling and Simulation (Africa-MS), 2008.
[13] Y. Qi, J. Klein-Seetharaman, Z. Bar-Joseph. "Random forest similarity for protein-protein interaction prediction from multiple sources," in Pacific Symposium on Biocomputing 10, pp. 531 - 542, 2005.
[14] T. K. Ho. "Random decision forests." ICDAR '95: Proceedings of the Third International Conference on Document Analysis and Recognition, Vol. 1, 1995.
[15] L. Breiman, A. Cutler. "Random forests." Department of Statistics, University of California Berkeley. http://www.stat.berkeley.edu/~breiman/RandomForests/cc_home.htm. Last accessed 12 August, 2008.
[16] J. R. Brence, D. E. Brown. "Improving the robust random forest regression algorithm." Systems and Information Engineering Technical Papers, Department of Systems and Information Engineering, University of Virginia, 2006. http://www.sys.virginia.edu/techreps/2006/sie06_0004.pdf. Last accessed 10 August, 2008.
[17] L. Breiman. "Random forests." Machine Learning, 45:5–32, Kluwer Academic Publishers, 2001.
[18] A. P. Engelbrecht. Computation Intelligence, an Introduction. John Wiley & Sons, Ltd, 2002.
[19] S. Haykin. Neural Networks: A Comprehensive Foundation. NY Macmillan. 1994.
[20] J-S. R. Jang, N. Gulley, "Fuzzy logic toolbox," The MathWorks Inc., 1997.
[21] A. Abraham, "Neuro-fuzzy systems: state-of-the-art modeling techniques," in Lecture Notes in Computer Science, Vol. 2084, pp. 269 - 276, Springer Verlag Germany, 2001.
[22] J-S. R. Jang, "Neurofuzzy modelling and control." Proceedings of the IEEE, Vol. 83, Issue 3, 1995.
[23] J-S. R. Jang, "Input selection for ANFIS learning." Proceedings of IEEE International Conference on Fuzzy Systems, 1998.
[24] E. E. E. Ali. "A proposed genetic algorithm selection method." King Saud University, ccis, 2006.
[25] H. Wong. "Genetic algorithms." Surprise 96 Journal, Imperial College of Science Technology and Medicine, 1996.
[26] A. M. S. Zalzala, P. J. Fleming. Genetic Algorithms in Engineering Systems. IET, 1997.
[27] T. Wang. "An interface to the random forest algorithm (version 3.3) written by Leo Breiman and Adele Cutler." StatLib—-Software and extensions for MATLAB. http://lib.stat.cmu.edu/matlab/. Last accessed 11 August, 2008.
[28] I.T. Nabney, C Bishop. "Netlab." http://www.ncrg.aston.ac.uk/netlab/. Last accessed 23 February, 2008.
[29] J. Joines. "Genetic algorithm optimization toolbox (GAOT) for Matlab 5." http://www.ise.ncsu.edu/mirage/GAToolBox/gaot/. Last accessed 10 April 2008.
[30] Y. Yuan, R. Lorenzo, C. Andrea. "On early stopping in gradient descent learning," in Constructive Approximation, Vol. 26, No. 2, pp. 289-315, 2007.
[31] J. Mistry, F.V. Nelwamondo, T. Marwala. "Investigation of autoencoder neural network accuracy for computational intelligence methods to estimate missing data." The IASTED International Conference on Modelling and Simulation, 2008.
[32] E. J. Wegman, J. L. Solka, "The curse of dimensionality," from Short Course in Data Mining. Eighth U.S. Conference on Applied Statistics, 2002.
[33] J. D. Gibbons, S. Chakraborti. Nonparametric Statistical Inference, pp. 2, 8, 19, 135, 144, CRC Press, Fourth Edition, 2003.
[34] P. C. Mahalabonis. "Generalized distance in statistics." Proceedings of the National Institute of Science of India 12, 1936.
[35] M. Vazirgiannis, M. Halkidi, D. Gunopulos. Uncertainty Handling and Quality Assessment in Data Mining, pp. 77, 78, Springer, 2003.
[36] N. Ye. The Handbook of Data Mining, pp. 431, Routledge Taylor & Francis Group, First Edition, 2003.
[37] F. C. Pampel. Logistic Regression - A Primer, pp. 1 - 4, SAGE, 2000.